\documentstyle[epsfig]{aipproc}
\newcommand{\beq}{\begin{equation}}
\newcommand{\eeq}{\end{equation}}
\newcommand{\bea}{\begin{eqnarray}}
\newcommand{\eea}{\end{eqnarray}}
\newcommand{\apj}{{\it Astrophys. J.} }

\newcommand{\rem}[1]{ }
\begin{document}
\title{Physics of Collisionless GRB Shocks and Their Radiation Properties}

\author{Mikhail V. Medvedev }
\address{CITA,University of Toronto, Toronto, Ontario, M5S 3H8, Canada\\
Harvard-Smithsonian Center for Astrophysics, 60 Garden Street,
Cambridge, MA 02138}

\maketitle

\begin{abstract}

We present a theory of ultrarelativistic collisionless shocks 
based on the relativistic kinetic two-stream instability. We
demonstrate that the shock front is unstable to the generation
of small-scale, randomly tangled magnetic fields. These fields
are strong enough to scatter the energetic incoming (in the shock
frame) protons and electrons over pitch angle and, therefore, 
to convert their kinetic energy of bulk motion into heat with 
very high efficiency. This validates the use of MHD approximation 
and the shock jump conditions in particular. The effective collisions 
are also necessary for the diffusive Fermi acceleration of electrons 
to operate and produce an observed power-law. Finally, these strong
(sub-equipartition) magnetic fields are also required for the efficient 
synchrotron-type radiation emission from the shocks.

The predicted magnetic fields have an impact on polarization properties
of the observed radiation (e.g., a linear polarization from a jet-like 
ejecta and polarization scintillations in radio for a spherical one)
and on its spectrum.
We present an analytical theory of jitter radiation, which is emitted 
when the magnetic field is correlated on scales smaller then the gyration 
(Larmor) radius of the accelerated electrons. The spectral power of jitter 
radiation is described by a sharply broken power-law: $P(\nu)\propto\nu^1$ 
for $\nu<\nu_j$ and $P(\nu)\propto\nu^{-(p-1)/2}$ for $\nu>\nu_j$, 
where $p$ is the electron power-law index and $\nu_j$ is the jitter break, 
which is independent of the magnetic field strength but depends on the shock
energetics and kinematics. Finally, we present a composite jitter+synchrotron 
model of GRB $\gamma$-ray emission from internal shocks which is capable 
of resolving many puzzles of GRB spectra, such as the violation of the 
``line of death'', sharp spectral breaks, and  multiple
spectral components seen in some bursts (good examples are
GRB910503, GRB910402, etc.). We stress that simultaneous detection of 
both spectral components opens a way to a precise diagnostics of the 
conditions in GRB shocks. We also discuss the relation of our
results to other systems, such as internal shocks in blazars,
radio lobes, and supernova shocks.
\end{abstract}

\section*{The structure of collisionless shocks}

The conventional paradigm of GRBs assumes optically thin synchrotron 
radiation from ultra-relativistic shocks where the radiation
is produced by Fermi-accelerated electrons moving in strong,
nearly equipartition magnetic fields. This purely phenomenological
model contains several serious assumptions which require justification:
(i) standard hydrodynamic shock physics must be valid for these
highly collisionless shocks, i.e., one needs effective collisions;
(ii) magnetic fields must be generated {\em in situ} much faster than 
the dynamical time; (iii) acceleration of electrons requires 
multiple scatterings (i.e., effective collisions) in the shock.

These problems have successfully been resolved \cite{ML99,Mtexas00}.
It was shown that magnetic fields are naturally produced via the 
relativistic two-stream instability operating at the shock front.
In essence, this work \cite{ML99} makes the bridge between the theories
non-relativistic collisionless shocks \cite{MS63} (those observed in the 
interplanetary space were studied {\em in situ} in great details by many 
satellites, and their ultra-relativistic counterparts. Here we briefly 
describe the main results.
\begin{itemize}
\item 
The two-stream instability operates in both internal and external shocks.
The field is produced by both the electrons and protons.
\item
The generated magnetic field is randomly oriented in space, but always lies
in the plane of the shock front.
\item
The characteristic $e$-folding time in the shock frame for the instability is
$\tau\sim{\gamma_{\rm sh}^{1/2}}/{\omega_{\rm p}}$ (where 
$\gamma_{\rm sh}$ is the shock Lorentz factor) which is $\sim10^{-7}~{\rm s}$ 
for internal shocks and $10^{-4}~{\rm s}$ for external shocks.
This time is much shorter than the dynamical time of GRB fireballs.
\item
The characteristic coherence scale of the generated magnetic field is of
the order of the relativistic skin depth
$\lambda\sim{c\bar\gamma^{1/2}}/{\omega_{\rm p}}$
(where $\bar\gamma$ is the mean thermal Lorentz factor of particles), i.e. 
$\sim10^3~{\rm cm}$ for internal shocks and $\sim10^5~{\rm cm}$ for external 
shocks. This scale is much smaller than the spatial scale of the source.
\item
The instability converts a large fraction of the kinetic energy of
particles into magnetic energy, hence 
$[{B^2/8\pi}]/[{mc^2n(\bar\gamma-1)}]=\eta\sim 10\%$.
This agrees well with direct particle simulations.
\item
Random fields scatter particles over pitch-angle and, thus, provide effective 
collisions. Therefore MHD approximation works well for the shocks.
The magnetic fields communicate the momentum and pressure of the outflowing 
fireball plasma to the ambient medium and define the shock boundary.
\item
The instability isotropizes and heats the electrons and protons.
Moreover, effective collisions will diffusively further accelerate 
the electrons to higher energies.
\end{itemize}

\section*{Radiation from shocks}

Since the geometry of magnetic fields is not entirely random, it affects
the observed properties of radiation. In particular, polarization of 
radiation will always be radial. Therefore, one expects non-vanishing
degree of polarization observed at any wavelength for a non-spherically 
symmetric explosion, e.g., a jetted geometry. An optical transient of 
GRB990510 shows linear polarization of the degree $\sim1-2\%$. For a 
spherically  symmetric case, radio scintillations may reveal the 
polarization map of the source (afterglow) \cite{M00}.

\begin{figure}[b!] 
\centerline{\epsfig{file=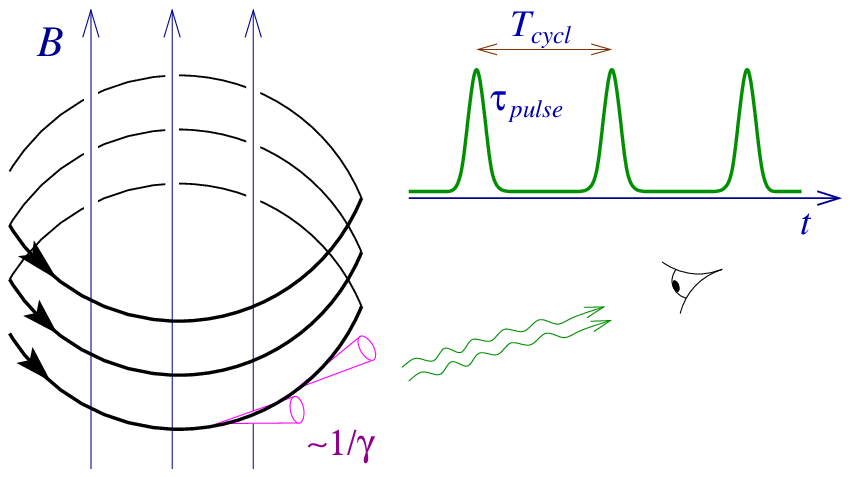,height=1.9in}
\epsfig{file=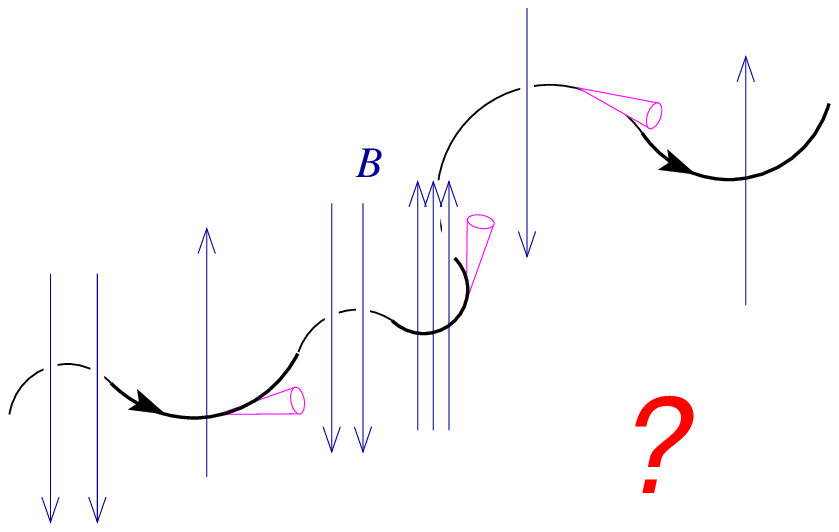,height=1.9in}}
\vspace{10pt}
\caption{Illustration of the process of radiation in
homogeneous and inhomogeneous fields.}
\label{fig1}
\end{figure}

The small-scale nature of magnetic fields affects the radiation process
as well. In fact, it breaks the conventional paradigm of synchrotron 
nature of the radiation from shocks.
The magnetic field produced in GRB shocks randomly fluctuates on a very
small scale of roughly the relativistic skin depth, which is much smaller 
than the Larmor radius of the ultra-relativistic emitting electron.
Therefore, the electron trajectories are not helical, as they would be 
in a homogeneous field, as in Figure \ref{fig1}. Thus, the theory of 
synchrotron radiation derived for homogeneous fields is not applicable
and the spectrum of the emergent radiation is different. Such a situation 
has never been considered in the astrophysical literature.

If the magnetic field is randomly tangled and the correlation length is less 
then a Larmor radius of an emitting electron, then the electron experiences
random deflections as it moves through the field. Its trajectory is,
in general, stochastic. This is similar to a collisional motion of 
an electron in a medium. Bremsstrahlung quanta are emitted in every collision.
Unlike the bremsstrahlung case, here ``collisions'' are due to small-scale 
inhomogeneities of the magnetic field rather than due to electrostatic fields 
of other charged particles. Since the Lorentz force depends on particle's
velocity, the emergent spectrum will be somewhat different from pure 
bremsstrahlung.
There is also an alternative physical interpretation of the process.
For an ultrarelativistic electron, the method of virtual quanta applies. 
In the rest frame of the electron, the magnetic field inhomogeneity
with wavenumber $k\sim1/\lambda$ is transformed into a transverse pulse 
of electromagnetic radiation with frequency $kc$. This radiation is then 
Compton scattered by the electron to produce observed radiation with 
frequency $\sim\gamma^2kc$ in the lab frame.  

Keeping this general physical picture in mind, we now analyze the problem
in more details.
Let's  consider a nonuniform random magnetic field with a typical 
correlation scale $\lambda_B$, the Larmor radius of the electron,
$\rho_e=\gamma m_ec^2/eB_\bot$ is less or comparable  comparable to 
$\lambda_B$. The emerging spectrum depends on the relation between the
particle's deflection angle, $\alpha$, and the beaming angle, $\Delta\theta$.
For ultrarelativistic 
particles and small deflection angles, the latter is estimated as follows. 
The particle's momentum is $p\sim\gamma m_ec$. The change in the 
perpendicular momentum due to the Lorentz force acting on the particle during
the transit time $t\sim\lambda_B/c$ is $p_\bot\sim F_Lt\sim eB_\bot\lambda_B/c$.
The angle $\alpha$ is then $\alpha\sim p_\bot/p\sim 
eB_\bot\lambda_B/\gamma m_ec^2$. We now define the deflection-to-beaming ratio
as follows,
\beq
\delta\equiv\frac{\gamma}{k_B\rho_e}\sim\gamma\frac{\lambda_B}{\rho_e}
\sim\frac{\alpha}{\Delta\theta}\sim\frac{eB_\bot\lambda_B}{m_ec^2}.
\eeq
It is interesting to note that this ratio is independent of particle's 
energy (i.e., of $\gamma$) and is determined by $B$ and $\lambda_B$. 

\begin{figure}[b!] 
\centerline{\epsfig{file=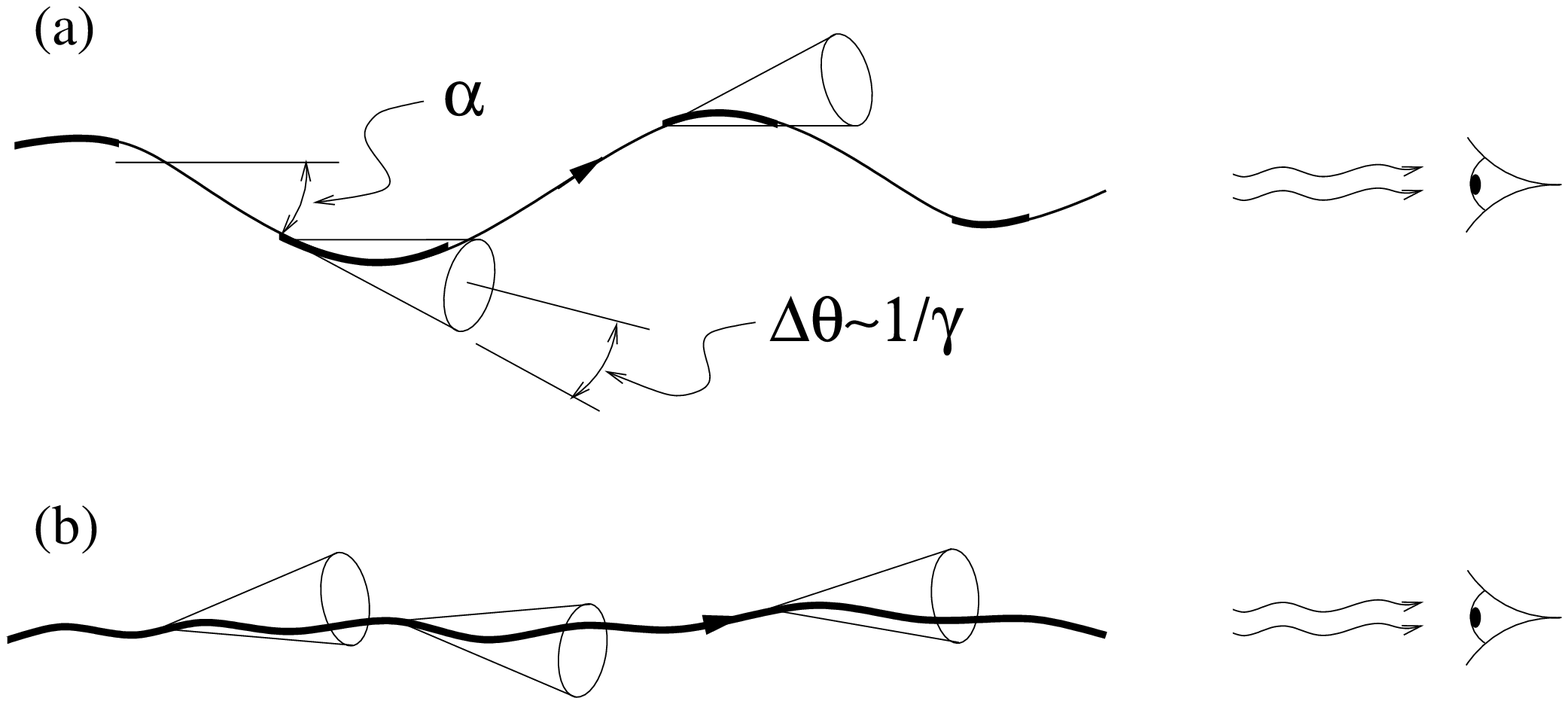,height=1.3in}~~~~
\epsfig{file=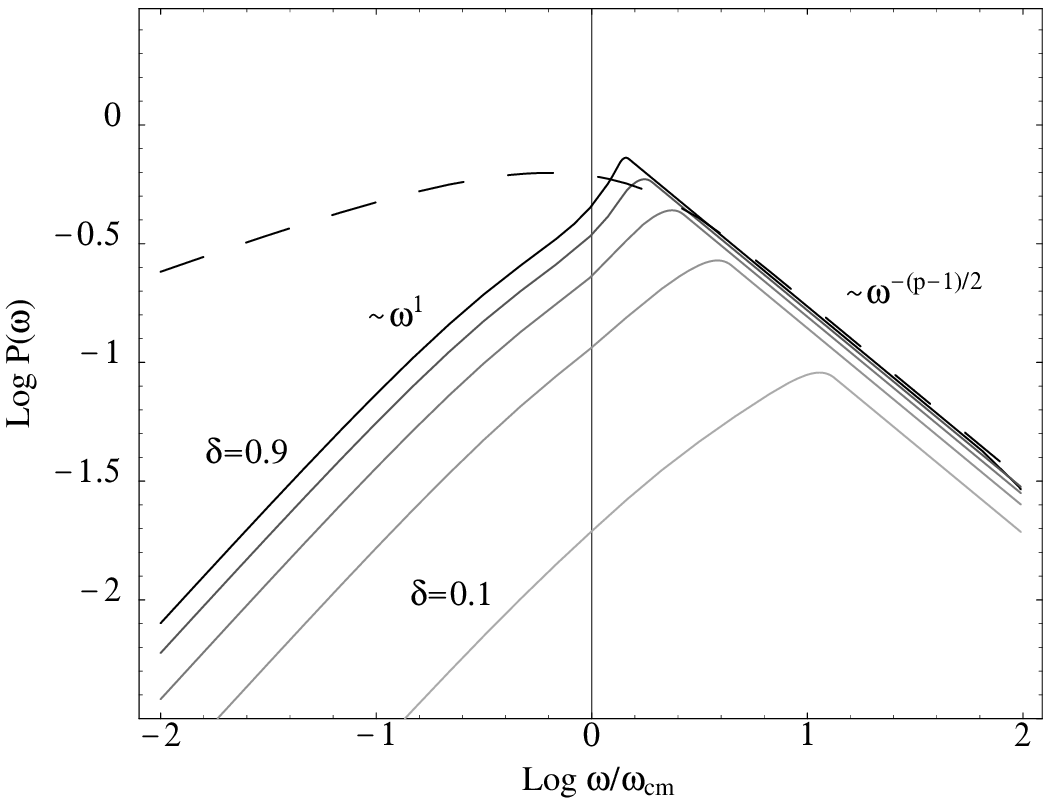,height=2.1in}}
\vspace{10pt}
\caption{(a,b) The emission process in $\delta\sim\alpha/\Delta\theta\gg1$ 
and $\delta\ll1$ regimes. (c) Spectral power of jitter radiation for various 
$\delta$; synchrotron spectrum is shown by dashed curve.}
\label{fig2}
\end{figure}

There are two limiting cases, as in Figure \ref{fig2}a,b.
First, $\delta\sim\alpha/\Delta\theta\gg1$; an observer sees radiation coming 
from short segments (``patches'') of the electron's trajectory, the
magnetic field is almost uniform but it varies from patch to patch.
The radiation is pulsed with a typical duration $\tau_p\sim 1/\omega_c$ as 
for pure synchrotron. The ensemble-averaged spectrum completely identical to 
synchrotron radiation from large-scale weakly inhomogeneous magnetic fields.
Second, $\delta\sim\alpha/\Delta\theta\ll1$; the particle moves along the 
line of sight almost straight and experiences high-frequency jittering in 
the perpendicular direction due to the random Lorentz force. The emergent 
spectrum is determined by random accelerations of the particle. 

Spectra of jitter radiation were calculated in Ref. \cite{M00}. 
They are well approximated by the sharply broken power-law, as is
seen from Figure \ref{fig2}c. The jitter break frequency is
\beq
\omega_{jm}=2^{7/4}
\gamma_{\rm sh}\gamma_{\rm int}\gamma_{\rm min}^2\bar\gamma_e^{-1/2}
\omega_{pe}\propto n_e^{1/2},
\eeq
where 
$\gamma_{\rm sh},\ \gamma_{\rm min},\ \gamma_{\rm int},\ \bar\gamma_e$
are the Lorentz factors of the ejecta, internal shock, electron power-law 
cutoff, and the mean thermal $\gamma$-factor of electron in front of the shock.
Note, this break frequency is independent of the magnetic field strength; 
instead it directly ``measures'' the density of particles in the shock.
Below and above the break, the photon spectra scale as ($p$ is the electron
power-law index)
\beq
F(\omega<\omega_{jm})\propto \omega^0, \qquad
F(\omega>\omega_{jm})\propto \omega^{-(p+1)/2},
\eeq
that is the spectrum is harder than synchrotron ($\propto\omega^{-2/3}$) 
at low frequencies.

As was mentioned earlier, the magnetic field is produced by both the 
electrons and the protons. However, only the electron-produced field 
is small-scale enough to produce jitter radiation, whereas for the 
proton-produced field $\delta>1$ and synchrotron spectrum is expected.
The ratio of the jitter and synchrotron break frequencies and peak 
fluxes completely determine both free parameter of the model:
the small-to-large scale field strength ratio and 
the deflection-to-beaming ratio,
\beq
\frac{\omega_{jm}}{\omega_{sm}}
\simeq\frac{2}{3}\frac{\bar B_{SS}}{\bar B_{LS}}\,\delta^{-1} ,\qquad
\frac{F_{J,{\rm max}}}{F_{S,{\rm max}}}\simeq\delta^2.
\eeq

This jitter+synchrotron model explains perfectly well the diversity
of time-resolved GRB $\gamma$-ray spectra and resolves several 
long-standing puzzles, namely
\begin{itemize}
\item
the violation of the constraint on the low-energy spectral index called
the synchrotron ``line of death'' in about a third of BATSE and BSAX 
bursts \cite{Preece+00};
\item
the sharp spectral break at the peak frequency seen in some bursts which 
is inconsistent with the broad synchrotron bump \cite{Pelaez+94};
\item 
the evidence for two spectral sub-components seen in some GRBs \cite{Barat+00};
\item
possible existence of emission features (``GRB lines'') seen in
few bursts \cite{Schaefer+98}.
\end{itemize}
All this strongly supports
that (i) the proposed jitter radiation mechanism operates in astrophysical 
objects and (ii) the magnetic field is generated in shocks by the two-stream 
instability. In general, the detection of both spectral components in GRB 
spectra would be a powerful and precise tool to investigate the properties of 
cosmological fireballs.


\begin{references}
%
\bibitem{Pi99}	
%
Piran, T., 
Phys. Rep. {\bf 314}, 575 (1999)
%
\bibitem{ML99}		
Medvedev, M. V., and Loeb, A., \apj {\bf 526}, 697 (1999)
%
\bibitem{Mtexas00}
Medvedev, M. V., in Proceedings of the 20th Texas Symposium,
eds. J. C. Wheeler and H. Martel (2001)
%
\bibitem{MS63}	
Moiseev, S. S., and  Sagdeev, R. Z., {\it J. Nucl. Energy C} {\bf 5}, 43 (1963)
%
\bibitem{M00}		
Medvedev, M. V., \apj {\bf 540}, 704 (2000)
%
\bibitem{Preece+00}	
Preece, R. D., et al., \apj Suppl. {\bf 126}, 19 (2000)
%
\bibitem{Pelaez+94}	
Pelaez, F., et al., \apj Suppl. {\bf 92}, 651 (1994)
%
\bibitem{Barat+00}	
Barat, C., et al., \apj {\bf 538}, 152 (2000)
%
\bibitem{Schaefer+98}	
Schaefer, B. E., et al., \apj {\bf 492}, 696 (1998)
%
\end{references}
\end{document}